\documentclass[runningheads]{llncs}
\usepackage{graphicx}
\usepackage{hyperref}
\usepackage{xcolor}
\usepackage{amsfonts} 
\usepackage{multirow}
\begin{document}
\title{Learning Sequential Information in Task-based fMRI for Synthetic Data Augmentation} 
\titlerunning{tfMRI Synthetic Data Augmentation}
\author{Jiyao Wang\inst{1,}\thanks{Corresponding author: Email: jiyao.wang@yale.edu}\and
Nicha C. Dvornek\inst{1,2} \and
Lawrence H. Staib\inst{1,2} \and
James S. Duncan\inst{1,2,3,4}}
\authorrunning{J. Wang et al.}
%
\institute{Biomedical Engineering, Yale University, New Haven, CT 06511, USA \and
Radiology \& Biomedical Imaging, Yale School of Medicine, New Haven, CT 06511, USA \and
Electrical Engineering, Yale University, New Haven, CT 06511, USA \and
Statistics \& Data Science, Yale University New Haven, CT, 06511, USA}
\maketitle              
\begin{abstract}
Insufficiency of training data is a persistent issue in medical image analysis, especially for task-based functional magnetic resonance images (fMRI) with spatio-temporal imaging data acquired using specific cognitive tasks. In this paper, we propose an approach for generating synthetic fMRI sequences that can then be used to create augmented training datasets in downstream learning tasks. To synthesize high-resolution task-specific fMRI, we adapt the $\alpha$-GAN structure, leveraging advantages of both GAN and variational autoencoder models, and propose different alternatives in aggregating temporal information. The synthetic images are evaluated from multiple perspectives including visualizations and an autism spectrum disorder (ASD) classification task. The results show that the synthetic task-based fMRI can provide effective data augmentation in learning the ASD classification task. 

\end{abstract}

\section{Introduction}
Synthetic data augmentation is a frequently used method in training machine learning models when training data is insufficient \cite{gan_aug1,gan_aug2,gan_aug3,vae_aug,fMRI_aug,cardiac_aug}. Although its usefulness has been demonstrated in a variety of fields related to medical imaging, most use cases are targeted towards either 2D \cite{gan_aug1,gan_aug2,vae_aug} or 3D images \cite{gan_aug3} that contain only spatial information. Only a few works explore synthetic augmentation of 4D imaging data including temporal information 
\cite{cardiac_aug,fMRI_aug}, but fMRI is still synthesized as an individual 3D frame \cite{fMRI_aug}. In this paper, we focus on augmenting the full spatio-temporal fMRI sequences from a task-based brain fMRI dataset acquired under an autism spectrum disorder (ASD) study. We show that augmenting the task-specific fMRI using an image synthesis model improves model robustness in a baseline spatio-temporal fMRI classification task. Moreover, the ability to generate synthetic fMRI data will enable fairer comparisons of different classes of models that can be trained on the same augmented dataset, removing bias introduced by model-specific data augmentation methods.

The generative adversarial network (GAN) \cite{gan} and variational autoencoder (VAE) \cite{vae} are two popular models in image synthesis. While GANs usually suffer from disadvantages such as mode collapse and the checker-board artifact, image resolution is a challenge for VAEs. The $\alpha$-GAN \cite{agan,gan_aug3} architecture is a promising alternative. It modifies the GAN architecture to include auto-encoding and embedding distribution features of VAE. For our experiment, we implement an $\alpha$-GAN for 4D input data to synthesize target fMRI. 

In previous years, recurrent neural network structures such as long short-term memory (LSTM) \cite{lstm} were frequently applied when learning sequential data. Recently, transformer structures \cite{att,vswin,bert}, including the application of the Swin transformer in video learning \cite{vswin}, provide the possibility to capture long-term information in spatio-temporal data using an attention approach. Moreover, the design of the BERT \cite{bert} model highlights a potential advantage of incorporating bidirectional information in capturing sequential data. In our implementation of $\alpha$-GAN, we extract spatial features from the brain using 3D convolution operations and experiment with alternatives in handling sequential spatial features including 1D convolution, LSTM, and attention. 

In summary, the contributions of this work are as follows:
\begin{itemize}
    \item We adapt the $\alpha$-GAN architecture to synthesize 4D task-based fMRI data, which is to our knowledge the first to synthesize the entire spatio-temporal sequence of task-based fMRI.
    \item We investigate different approaches for performing temporal aggregation within the $\alpha$-GAN network.
    \item We assess the effectiveness of fMRI image generation through quantitative analysis on brain regions related to the fMRI task, sample visualizations, and downstream use of the synthetic data in an ASD classification task.
\end{itemize}

\section{Model Architecture}
\begin{figure}
\includegraphics[width=\textwidth]{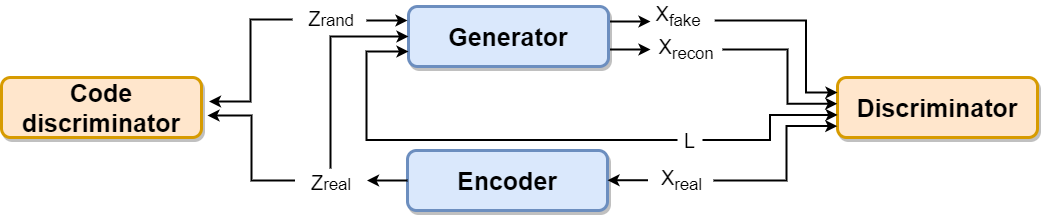}
\caption{$\alpha$-GAN model structure} \label{fig1}
\end{figure}

Following the design in Rosca et al.\ \cite{agan}, our $\alpha$-GAN model for fMRI data synthesis has four components: an encoder, a generator, a discriminator, and a code discriminator. For our application, the encoder maps a sequence of 3D volumes $X=(x_1,x_2,\ldots,x_T)$ into a compact vector embedding $z$. Given an embedding $z$ and a class label $L$, the generator generates a 4D output $X$. The discriminator classifies input $X$ between real or synthetic. The code discriminator classifies $z$ as generated from real $X$ or from a random standard normal distribution (Fig.~\ref{fig1}).  

Compared to a typical GAN architecture consisting of only generator and discriminator, the $\alpha$-GAN model has two more components. The encoder component forms an auto-encoding structure with the generator, allowing us to utilize the reconstruction loss between the real image input to the encoder and the reconstructed output from the generator. This is especially beneficial for complex high-dimensional input data, providing the generator useful gradient information in addition to the adversarial feedback from the discriminator. It also allows us to pretrain the encoder-generator pair as an autoencoder. Meanwhile, the code discriminator component encourages the encoder-calculated embedding from real images to be similar to the embedding generated from a standard normal distribution, which is similar to the design of a VAE. 
Theoretically, the $\alpha$-GAN model generates more stable variations in synthetic images than a typical GAN.
In practice, we find that the $\alpha$-GAN model also considerably improves the resolution and fineness of details in synthetic images. 

\begin{figure}
\includegraphics[width=\textwidth]{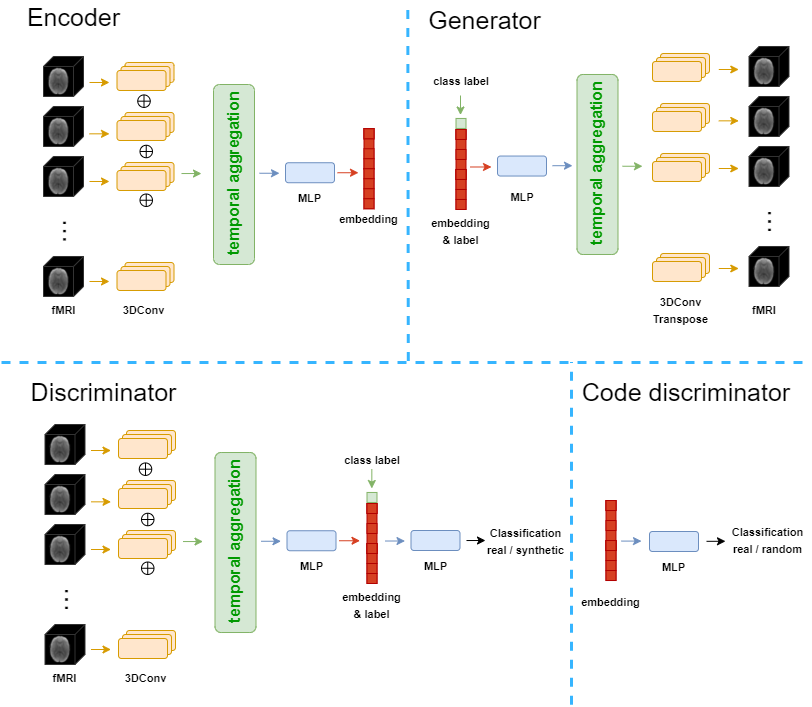}
\caption{$\alpha$-GAN architecture diagrams for all components described above (encoder, generator, discriminator, and code discriminator)} \label{fig2}
\end{figure}

As shown in Fig.~\ref{fig2}, we design the encoder and discriminator components in our model to process spatio-temporal information from sequential frames of fMRI. We first utilize 3D convolution to extract spatial features from each frame. Spatial features across frames are then processed by a temporal aggregation module. For the discriminator, an additional multilayer perceptron (MLP) module is included to produce the classification output. The generator component is an inverse of the encoder taking the image embedding and class label as input. Finally, the code discriminator is another MLP for classification.

\begin{figure}
\includegraphics[width=\textwidth]{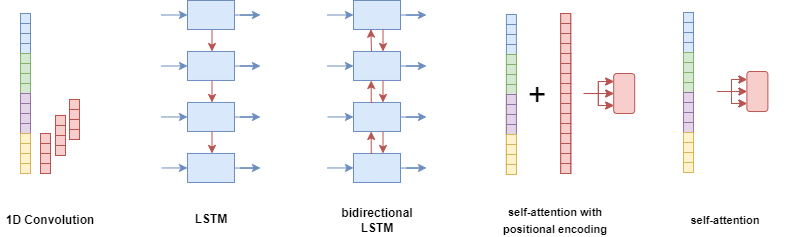}
\caption{Alternatives in processing temporal information applied in temporal aggregation modules of figure above} \label{fig3}
\end{figure}

For the extracted temporal information, we experiment with alternatives including 1D convolution, LSTM, bidirectional LSTM, self-attention with positional encoding, and self-attention without positional encoding (Fig.~\ref{fig3}).Theoretically, 1D convolution learns from a limited temporal kernel and shifts the same kernel along the entire sequence. It works better in capturing reoccurring local patterns. LSTM and bidirectional LSTM learn the temporal dependencies from one or both directions with a focus on remembering short-term dependencies for a long time.
The attention algorithm is good at capturing long-range dependencies. When the positional encoding is removed, learning depends only on the similarity between data without considering their temporal/spatial adjacency. 

\section{Data}
We use a 118-subject task-based fMRI dataset acquired under the "biopoint" biological motion perception task \cite{biopoint} designed to highlight deficits in motion perception in children with ASD. Subjects include 75 ASD children and 43 age-and-IQ-matched healthy controls. The data collection and study was approved by the Institutional Review Board (IRB) at Anonymous Institution (HIC \#1106008625). The obtained fMRI data is preprocessed using the pipeline described in Yang et al \cite{preprocess} with steps including: 1) motion correction, 2) interleaved slice timing correction, 3) BET brain extraction, 4) grand mean intensity normalization, 5) spatial smoothing, 6) high-pass temporal filtering. Each fMRI sequence contains 146 frames of $91\times109\times91$ 3D images with a frame interval of 2 seconds each. The voxel size is $3.2mm\times3.2mm\times3.2mm$. There are 12 task stimulation videos of biological and scrambled motion, which are well aligned between subjects during the data acquisition period and given in alternating sequence. We split the dataset into 70/15/15\% training/validation/test data, resulting in 72/23/23 subjects in each subset.

\section{Training}

During training, we apply a two-stage training scheme for the $\alpha$-GAN model described above. In the first pre-training stage, the encoder and generator components are trained briefly (around 20 epochs) as an autoencoder network towards a minimum mean squared error (MSE) on 4D fMRI image reconstruction. Learned weights for both components are loaded in the second training stage to provide stable reconstruction performance at initialization. In the second training stage, training of the $\alpha$-GAN model takes 3 steps including training the encoder-generator pair, discriminator, and code discriminator respectively. Let $X_{real}, X_{recon}, X_{fake}$ denote the input fMRI images, reconstructed fMRI images, and synthesized fMRI images from random embedding. $z_{real}$ and $z_{rand}$ denote the embedding generated from the encoder and a code sampled from the random standard normal distribution, respectively. The encoder $E$ and generator $G$ of our model are trained together to minimize a loss function consisting of 3 loss terms: 1) Mean absolute error (MAE) reconstruction loss between input image $X_{real}$ and reconstructed image $X_{recon}$; 2) Cross entropy (CE) loss optimizing the encoder-generator pair to generate $X_{recon}, X_{fake}$ that the discriminator classifies to be real images; 3) CE loss optimizing the encoder to generate $z_{real}$ that the code discriminator classifies to be an image embedding generated from a random standard normal distribution. Discriminator $D$ is trained to classify $X_{real}$ as 1, $X_{recon}$ and $X_{fake}$ as 0 using CE loss. Code discriminator $C$ is also trained using CE loss to classify $z_{real}$ as 1, $z_{rand}$ as 0. The losses are summarized below,
\begin{equation}
loss_{E, G}=\lambda ||x_{real}-x_{recon}||_1 -\log D(x_{recon})-\log D(x_{fake})-\log (1-C(z_{real}))
\end{equation}
\begin{equation}
loss_{D}=-\log D(x_{real}) -\log (1-D(x_{recon})) -\log (1-D(x_{fake}))
\end{equation}
\begin{equation}
loss_{C}=-\log (C(z_{real}))-\log (1-C(z_{rand}))
\end{equation}
where $x\in \mathbb{R}^{91\times109\times91\times146}$ and $z\in \mathbb{R}^{864}$. The models are implemented using PyTorch 1.10.2 \cite{pytorch} package and trained with the Adam optimizer under 100 epochs and a mini-batch of size 1. The learning rates for encoder-generator pair, discriminator, and code discriminator are 4, $1\times10^{-6}$, and $2\times10^{-5}$ respectively. There are four consecutive 3D convolution layers for the encoder with parameters: kernel size $=$ 16, 8, 4, 2, stride $=$ 2, 2, 2, 1, and dimension $=$ 4, 8, 16, 24. The generator is an inverse of the encoder using transpose convolution. The discriminator has three 3D convolution layers with parameters: kernel size $=$ 8, 4, 4, stride $=$ 4, 2, 1, and dimension $=$ 4, 8, 16. For the temporal aggregation methods, 1D convolution has two layers with kernel size of 8 and stride of 4. For LSTM, we use two layers of LSTM and half the feature dimensions when changing to bidirectional. For dot-product self-attention, we use one layer of attention with raster sequence positional encoding. Training each model takes approximately 40 hours on a single Nvidia A100 GPU. Comparison of real and generated image samples is shown in Supplementary Figure 1.

\section{Evaluation and Result}
First, we quantitatively analyze the similarity of fMRI signals between real data and similar-sized samples of synthetic data in three brain regions: right amygdala, fusiform gyrus, and ventromedial prefrontal cortex. These regions were identified in a previous ASD biopoint study \cite{biopoint} as showing salient signal changes between biological motion videos (BIO) versus scrambled motion videos (SCRAM). Ideally, the synthetic fMRI should show similar signal changes in these regions. We first use the AAL3 atlas \cite{AAL3} to obtain parcellations and average signals of all voxels in each region. Then, we extract fMRI sequences under SCRAM and BIO stimulation respectively and calculate the average Z-score for both sequences in Table~\ref{tab1}. We also perform unpaired, two-tailed t-tests between signals in BIO and SCRAM frames. The p-values are listed in Table~\ref{tab2}. The bold text in each column shows the regional pattern most similar to real fMRI. From the Z-score and t-test evaluation, the model using 1D convolution has signal most similar to real fMRI in the right amygdala and fusiform gyrus. The highest similarity in the ventromedial prefrontal cortex is achieved by the model using self-attention with positional encoding. Note that the 1D convolution model exaggerates the signal contrast between BIO and SCRAM sequences for fusiform gyrus and prefrontal cortex. Still, the 1D convolution model is the only variation that produces Z-scores with the same sign as the real fMRI for all brain regions.

\begin{table}
\caption{Average Z-score of BIO and SCRAM Sequences}\label{tab1}
\centering
\addtolength{\tabcolsep}{4pt} 
\resizebox{1.0\textwidth}{!}{
\begin{tabular}{l | c c | c c | c c }\hline
\multirow{2}{*}{Method} & \multicolumn{2}{c}{Right Amygdala} & \multicolumn{2}{|c|}{Fusiform Gyrus}& \multicolumn{2}{c}{Prefrontal Cortex}\\ \cline{2-7}
& SCRAM & BIO & SCRAM & BIO & SCRAM & BIO\\
\hline
\hline
Real fMRI & 0.140 & -0.144 & 0.165 & -0.170 & -0.073 & 0.075\\
\hline
1D Convolution & \textbf{0.137} & \textbf{-0.140} & \textbf{0.224} & \textbf{-0.230} & -0.219 & 0.225\\
LSTM & -0.047 & 0.049 & 0.072 & -0.073 & -0.061 & 0.063\\
Bidirectional LSTM & -0.089 & 0.091 & 0.029 & -0.030 & 0.015 & -0.015\\
Self-attention w/ PE & 0.072 & -0.074 & -0.076 & -0.078 & \textbf{-0.069} & \textbf{0.071}\\
Self-attention w/o PE & -0.002 & 0.002 & -0.002 & 0.002 & -0.002 & 0.002\\
\hline
\end{tabular}
}
Bold text shows regional pattern most similar to real fMRI
\end{table}

\begin{table}[t]
\caption{T-test p-value Between BIO and SCRAM Sequences}\label{tab2}
\centering
\addtolength{\tabcolsep}{4pt} 
\begin{tabular}{l | c | c | c }\hline
\multirow{2}{*}{Method} & Right Amygdala & Fusiform Gyrus& Prefrontal Cortex\\ \cline{2-4}
& p-value & p-value & p-value \\
\hline
\hline
Real fMRI & 0.089 & 0.043 & 0.374 \\
\hline
1D Convolution & \textbf{0.095} & \textbf{0.006} & 0.007 \\
LSTM & 0.565 & 0.385 & 0.858 \\
Bidirectional LSTM & 0.281 & 0.722 & 0.455 \\
Self-attention w/ PE & 0.383 & 0.357 & \textbf{0.402} \\
Self-attention w/o PE & 0.978 & 0.978 & 0.978 \\
\hline
\end{tabular}
Bold text shows regional pattern most similar to real fMRI
\end{table}

\begin{figure}
\includegraphics[width=\textwidth]{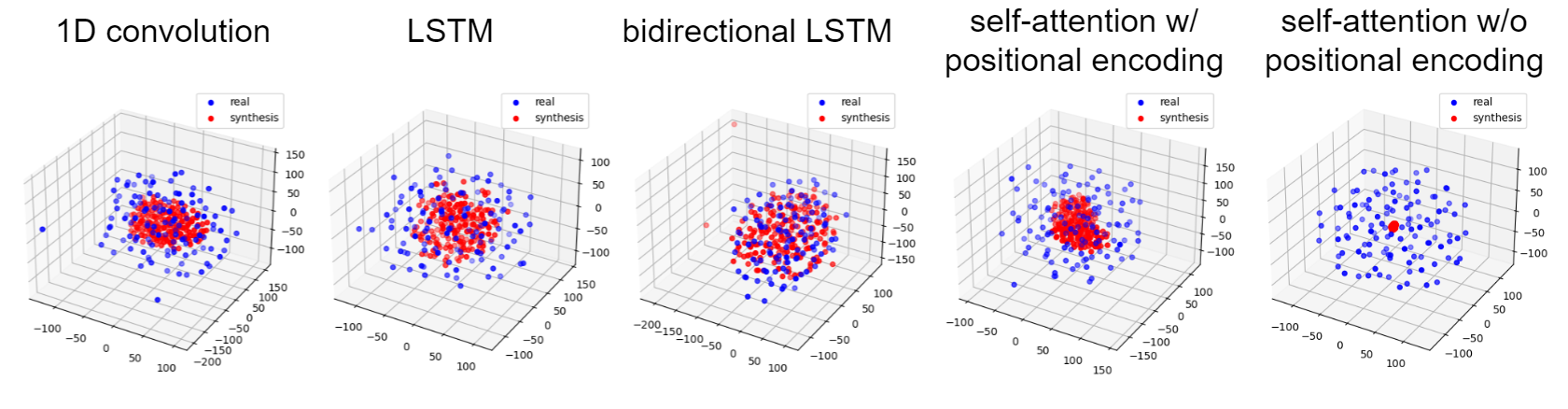}
\caption{Plots of tSNE projection. Each 4D fMRI is reduced to 100 dimensions by PCA and projected onto 3D by tSNE. Blue denotes real data, red denotes synthetic data.} \label{fig4}
\end{figure}

To compare the distributions of real vs. synthetic fMRI sequences, we perform a tSNE \cite{tSNE} visualization. We generate 200 synthetic fMRI consisting of 100 synthetic ASD subjects and 100 healthy control (HC) subjects for each temporal aggregation method. Then, we apply PCA and tSNE \cite{tSNE} to project the 118 real and 200 synthetic fMRI onto a 3-dimensional space. See Fig.~\ref{fig4}. All five alternatives for temporal aggregation generate synthetic data that have distribution centers similar to real fMRI in the spatio-temporal projection plots. However, considering the dispersion of data, the two plots of synthetic images generated using the attention algorithm have obviously less dispersion than the real fMRI, especially for the model trained without positional encoding. The 1D convolution result is better, while the two plots from the LSTM results have dispersion most similar to the distribution of real fMRI.

\begin{figure}
\includegraphics[width=\textwidth]{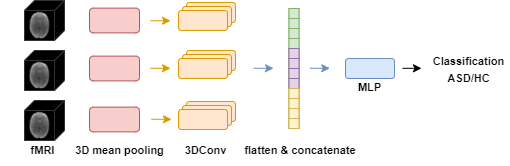}
\caption{Architecture of the classifier. Augmented 4D images are down-sampled spatially by mean pooling and passed to a convolutional network for an ASD classification task.} \label{fig5}
\end{figure}

In addition to evaluations via quantitative signal analysis and tSNE embedding, we also assess the utility of the synthetic data in augmenting training data for learning an ASD versus HC classification task. The architecture of the classifier is shown in Fig.~\ref{fig5}, which consists of 3D average pooling and 3D convolution operations to extract spatial features and an MLP module to calculate the classification output. The goal is to investigate the performance of synthetic fMRI for data augmentation. For classifier training, we use the 72-subject training subset of the fMRI dataset and augmented the training set to 792 samples by either adding random Gaussian noise ($\mu=0, \sigma=0.1$) or applying one of the five alternatives of the $\alpha$-GAN model. For synthesized images using each alternative, the number of subjects in ASD and HC groups are balanced. For model selection, we save the best model evaluated by lowest validation loss during training. The resulting performances on the testing set are listed in Table~\ref{tab3}. 

\begin{table}
\caption{Classifier Performances on Test Set }\label{tab3}
\centering
\addtolength{\tabcolsep}{4pt} 
\begin{tabular}{l|c c c c}
\hline
Method &  Testing CE Loss  &  Testing Acc (\%)  &   F1 Score  &  AUC\\
\hline
\hline
w/o augmentation & 0.609 & 69.6 & 0.759 & 0.795\\
Gaussian & 0.731 & 52.2 & 0.686 & 0.697\\
\hline
1D Convolution & 0.571 & \textbf{78.3} & \textbf{0.815} & \textbf{0.833}\\
LSTM & 0.618 & 73.9 & 0.769 & 0.758\\
Bidirectional LSTM & 0.613 & 73.9 & 0.720 & 0.765\\
Self-attention w/ PE & \textbf{0.505} & \textbf{78.3} & 0.800 & 0.814\\
Self-attention w/o PE & 0.634 & 69.6 & 0.741 & 0.735\\
\hline
\end{tabular}
\end{table}

There are two tasks for our $\alpha$-GAN model. Explicitly, we want to generate synthetic 4D fMRI that are similar to real images using the adversarial competition between generator and discriminator. Implicitly, as a variation of the conditional GAN model \cite{conditional-gan}, we expect the synthetic images to preserve the ASD versus HC class differences. Evaluated by the results above, the images generated using 1D convolution and self-attention with positional encoding approaches have the best performance on the implicit task. Meanwhile, both approaches show noticeable improvement compared to learning from the raw dataset without augmentation. 


\section{Conclusion}

Considering all the evaluations, 1D convolution produced the best overall performance. LSTM is usually considered a good choice for handling sequential information, but does not perform as well on our generation task.
Meanwhile, the experimental results of the attention models agree with the conclusion in \cite{compare-transformer-cnn} that non-pre-trained convolutional structures are competitive and usually outperform non-pre-trained attention algorithms. 
Furthermore, the performance across temporal aggregation methods also enables us to make hypotheses regarding task-based fMRI data. There are two intuitive perspectives of viewing task-based fMRI, stressing either the temporal dependencies between brain states or correspondence between brain signal and task stimulation. 
Our results might be an indication that the task-image-correspondence plays a more important role in explaining task-based fMRI than we expected. 

In recent years, various machine learning models have been applied to analyze fMRI data, including CNN, LSTM, and GNN \cite{cnn-brain,nicha-lstm,BrainGNN}. Comparing performance within one category of models is straightforward, but comparing between categories includes bias from using different model-dependent data augmentation methods. 
Our method to synthesize the fMRI sequence directly removes this bias, as the same augmented dataset can be used to train all models. In the future, we intend to expand our experiments to large public datasets and apply this method as data augmentation for analysis of other fMRI data. 

%
%
\bibliographystyle{splncs04}
\bibliography{fMRI_aug}

\begin{thebibliography}{10}
\providecommand{\url}[1]{\texttt{#1}}
\providecommand{\urlprefix}{URL }
\providecommand{\doi}[1]{https://doi.org/#1}

\bibitem{cardiac_aug}
Abbasi-Sureshjani, S., Amirrajab, S., Lorenz, C., Weese, J., Pluim, J.,
  Breeuwer, M.: 4d semantic cardiac magnetic resonance image synthesis on xcat
  anatomical model. In: Arbel, T., Ben~Ayed, I., de~Bruijne, M., Descoteaux,
  M., Lombaert, H., Pal, C. (eds.) Proceedings of the Third Conference on
  Medical Imaging with Deep Learning. Proceedings of Machine Learning Research,
  vol.~121, pp. 6--18. PMLR (06--08 Jul 2020),
  \url{https://proceedings.mlr.press/v121/abbasi-sureshjani20a.html}

\bibitem{bert}
Devlin, J., Chang, M., Lee, K., Toutanova, K.: {BERT:} pre-training of deep
  bidirectional transformers for language understanding. CoRR
  \textbf{abs/1810.04805} (2018), \url{http://arxiv.org/abs/1810.04805}

\bibitem{nicha-lstm}
Dvornek, N., Ventola, P., Pelphrey, K., Duncan, J.: Identifying autism from
  resting-state fmri using long short-term memory networks. In: Machine
  learning in medical imaging. MLMI (Workshop). vol. 10541, pp. 362--370 (09
  2017). \doi{10.1007/978-3-319-67389-9\_42}

\bibitem{gan_aug1}
Frid-Adar, M., Klang, E., Amitai, M., Goldberger, J., Greenspan, H.: Synthetic
  data augmentation using gan for improved liver lesion classification. In:
  2018 IEEE 15th International Symposium on Biomedical Imaging (ISBI 2018). pp.
  289--293 (2018). \doi{10.1109/ISBI.2018.8363576}

\bibitem{gan}
Goodfellow, I.J., Pouget-Abadie, J., Mirza, M., Xu, B., Warde-Farley, D.,
  Ozair, S., Courville, A., Bengio, Y.: Generative adversarial networks (2014).
  \doi{10.48550/ARXIV.1406.2661}, \url{https://arxiv.org/abs/1406.2661}

\bibitem{lstm}
Hochreiter, S., Schmidhuber, J.: Long short-term memory. Neural computation
  \textbf{9},  1735--80 (12 1997). \doi{10.1162/neco.1997.9.8.1735}

\bibitem{biopoint}
Kaiser, M.D., Hudac, C.M., Shultz, S., Lee, S.M., Cheung, C., Berken, A.M.,
  Deen, B., Pitskel, N.B., Sugrue, D.R., Voos, A.C., Saulnier, C.A., Ventola,
  P., Wolf, J.M., Klin, A., Wyk, B.C.V., Pelphrey, K.A.: Neural signatures of
  autism. Proceedings of the National Academy of Sciences  \textbf{107}(49),
  21223--21228 (2010). \doi{10.1073/pnas.1010412107},
  \url{https://www.pnas.org/doi/abs/10.1073/pnas.1010412107}

\bibitem{vae}
Kingma, D.P., Welling, M.: Auto-encoding variational bayes (2013).
  \doi{10.48550/ARXIV.1312.6114}, \url{https://arxiv.org/abs/1312.6114}

\bibitem{gan_aug3}
Kwon, G., Han, C., Kim, D.s.: Generation of 3d brain mri using auto-encoding
  generative adversarial networks. In: Shen, D., Liu, T., Peters, T.M., Staib,
  L.H., Essert, C., Zhou, S., Yap, P.T., Khan, A. (eds.) Medical Image
  Computing and Computer Assisted Intervention -- MICCAI 2019. pp. 118--126.
  Springer International Publishing, Cham (2019)

\bibitem{BrainGNN}
Li, X., Zhou, Y., Dvornek, N., Zhang, M., Gao, S., Zhuang, J., Scheinost, D.,
  Staib, L.H., Ventola, P., Duncan, J.S.: Braingnn: Interpretable brain graph
  neural network for fmri analysis. Medical Image Analysis  \textbf{74},
  102233 (2021). \doi{https://doi.org/10.1016/j.media.2021.102233},
  \url{https://www.sciencedirect.com/science/article/pii/S1361841521002784}

\bibitem{vswin}
Liu, Z., Ning, J., Cao, Y., Wei, Y., Zhang, Z., Lin, S., Hu, H.: Video swin
  transformer (2021). \doi{10.48550/ARXIV.2106.13230},
  \url{https://arxiv.org/abs/2106.13230}

\bibitem{tSNE}
van~der Maaten, L., Hinton, G.E.: Visualizing data using t-sne. Journal of
  Machine Learning Research  \textbf{9},  2579--2605 (2008)

\bibitem{vae_aug}
Madan, Y., Veetil, I.K., V, S., EA, G., KP, S.: Synthetic data augmentation of
  mri using generative variational autoencoder for parkinson's disease
  detection. In: Bhateja, V., Tang, J., Satapathy, S.C., Peer, P., Das, R.
  (eds.) Evolution in Computational Intelligence. pp. 171--178. Springer Nature
  Singapore, Singapore (2022)

\bibitem{conditional-gan}
Mirza, M., Osindero, S.: Conditional generative adversarial nets. CoRR
  \textbf{abs/1411.1784} (2014), \url{http://arxiv.org/abs/1411.1784}

\bibitem{pytorch}
Paszke, A., Gross, S., Chintala, S., Chanan, G., Yang, E., DeVito, Z., Lin, Z.,
  Desmaison, A., Antiga, L., Lerer, A.: Automatic differentiation in pytorch
  (2017)

\bibitem{cnn-brain}
Qureshi, M.N.I., Oh, J., Lee, B.: 3d-cnn based discrimination of schizophrenia
  using resting-state fmri. Artificial Intelligence in Medicine  \textbf{98},
  10--17 (2019). \doi{https://doi.org/10.1016/j.artmed.2019.06.003},
  \url{https://www.sciencedirect.com/science/article/pii/S0933365719301393}

\bibitem{AAL3}
Rolls, E.T., Huang, C.C., Lin, C.P., Feng, J., Joliot, M.: Automated anatomical
  labelling atlas 3. NeuroImage  \textbf{206},  116189 (2020).
  \doi{https://doi.org/10.1016/j.neuroimage.2019.116189},
  \url{https://www.sciencedirect.com/science/article/pii/S1053811919307803}

\bibitem{agan}
Rosca, M., Lakshminarayanan, B., Warde-Farley, D., Mohamed, S.: Variational
  approaches for auto-encoding generative adversarial networks (2017).
  \doi{10.48550/ARXIV.1706.04987}, \url{https://arxiv.org/abs/1706.04987}

\bibitem{compare-transformer-cnn}
Tay, Y., Dehghani, M., Gupta, J.P., Bahri, D., Aribandi, V., Qin, Z., Metzler,
  D.: Are pre-trained convolutions better than pre-trained transformers? CoRR
  \textbf{abs/2105.03322} (2021), \url{https://arxiv.org/abs/2105.03322}

\bibitem{att}
Vaswani, A., Shazeer, N., Parmar, N., Uszkoreit, J., Jones, L., Gomez, A.N.,
  Kaiser, L., Polosukhin, I.: Attention is all you need (2017).
  \doi{10.48550/ARXIV.1706.03762}, \url{https://arxiv.org/abs/1706.03762}

\bibitem{gan_aug2}
Waheed, A., Goyal, M., Gupta, D., Khanna, A., Al-Turjman, F., Pinheiro, P.R.:
  Covidgan: Data augmentation using auxiliary classifier gan for improved
  covid-19 detection. IEEE Access  \textbf{8},  91916--91923 (2020).
  \doi{10.1109/ACCESS.2020.2994762}

\bibitem{preprocess}
Yang, D., Pelphrey, K.A., Sukhodolsky, D.G., Crowley, M.J., Dayan, E., Dvornek,
  N.C., Venkataraman, A., Duncan, J., Staib, L., Ventola, P., et~al.: Brain
  responses to biological motion predict treatment outcome in young children
  with autism. Translational Psychiatry  \textbf{6}(11) (2016).
  \doi{10.1038/tp.2016.213}

\bibitem{fMRI_aug}
Zhuang, P., Schwing, A.G., Koyejo, O.: Fmri data augmentation via synthesis.
  In: 2019 IEEE 16th International Symposium on Biomedical Imaging (ISBI 2019).
  pp. 1783--1787 (2019). \doi{10.1109/ISBI.2019.8759585}

\end{thebibliography}
\end{document}